\documentclass[11pt]{article}
\usepackage{acl2014}
\usepackage{times}
\usepackage{amsmath}
\usepackage{url}
\usepackage{latexsym}

\title{Who negatively influences me? Formalizing diffusion dynamics of negative exposure leading to student attrition in MOOCs \\{\small LTI Student Research Symposium 2014, Carnegie Mellon University, USA}}

\author{Tanmay Sinha \\
  Language Technologies Institute\\
  Carnegie Mellon University, Pittsburgh PA 15213\\
  {\tt tanmays@andrew.cmu.edu} \\}

\begin{document}
\maketitle
\begin{abstract}
In this work, we explain the underlying interaction mechanisms which govern students' influence on each other in Massive Open Online Courses (MOOCs).  Specifically, we outline different ways in which students can be negatively exposed to their peers on MOOC forums and discuss a simple formulation of learning network diffusion, which formalizes the essence of how such an influence spreads and can potentially lead to student attrition over time. We also view the limitations of our student modeling in the light of real world MOOC behavior and consequently suggest ways of extending the diffusion model to handle more complex assumptions. Such an understanding is very beneficial for MOOC designers and instructors to create a conducive learning environment that supports students' growth and increases their engagement in the course.
\end{abstract}

\section{Introduction}
Massive Open Online Courses (MOOCs) have generated a great deal of elation for their potential to provide students with the autonomy of grappling with the course instruction at their own understanding pace and connecting with millions of diverse learners from all over the world. However, despite having such a tremendous upper hand over traditional classroom and online learning settings, many rich learning activities that have been empirically validated with small classes seem, at first glance, difficult to scale up to thousands of learners. A pressing concern is the potential failure of MOOCs to produce a conducive learning environment which sustains collaboration among differently motivated learners and encourages them to stay in the course. Recent studies substantiate that dropout rate in MOOCs is more than 90\% \cite{Belanger:13,Schmidt:13,Yang:13}. Moreover, less than 5\% \cite{Huang:14} of students actively participate in MOOC discussion forums, which are central to students' collaboration, discussions on course related topics, exchange of ideas and information in these interactive learning networks \cite{Mason:12}.

While there is increasing focus on understanding MOOC discussion forum activities by studying a)thread starting, posting activity and social positioning in post reply discussion networks \cite{SinhaB:14,Yang:13}, b)engagement, motivation level and sentiment of posts \cite{WenA:14,WenB:14} for individual students, relatively little analysis has been done on examining how peer influence affects students' behavior, particularly in a manner that impedes progress and affects their decision to stay in the course. Though a related work \cite{Yang:14} investigates role of relational bonds in keeping students engaged, it fails to explain the reason behind how relationship loss (quantified by factors such as similar cohort membership, reply interaction, common thread participation, community connection and similar topic distribution in posts), diffuses through the MOOC learning network over time.

Moreover, Coursera MOOC datasets that we use in our work \cite{SinhaC:14} reveal that students have six to eight time greater instances of forum and thread viewing than explicit posting, commenting or thread starting activities. This might be explained by the fact that students have specific information needs to fulfil and so they might not actually need to create new content (analogous to the manner in which very few people actually need to ask generic questions on websites such as Stack Overflow, because they have already been answered). However, an alternate and a more plausible explanation is that because the MOOC forums are flooded with poor quality posts, students are less excited and find it less productive to actually engage in discussions with peers. This line of reasoning goes more in sync with the fact that very low active forum participation has been recorded for the first generation of MOOC offerings.

In light of these developments, it is important to explain the intricacies of students' negative or positive effect on each other in the MOOC over time. Existing literature reveals that peers influence attitudes and behaviors by acting as role models \cite{Kaplan:87}, reinforcing deviant behavior \cite{Esb:98}, and developing mutually influential norms that promote continued deviant behavior \cite{Dishion:94,Loe:87}. Thus, to orchestrate MOOC instruction, it would be helpful if we determine factors that might intercept MOOCs from turning into a healthy learning community, where students not only socialize but also mutually benefit from the forum participation of each other. The mutual participation involves discussions and crisp technical help on course related topics.

Thus, in this work, we operationalize the diffusion dynamics of negative exposure based on forum interaction footprint of students. For MOOC instructors, the focal utility is thus two-fold. 
\begin{itemize}
\item Firstly, they can have apriori information on which wave of students are likely to be influenced in future, and can consequently target interventions for these learners, to positively motivate and pull them back in remaining time steps before they actually dropout.
\item Secondly, knowledge about students who spread such a negative influence in MOOCs is extremely useful to allocate extra resources to counsel them or provide recommendations on more fruitful and engrossing ways to engage with the course. 
\end{itemize}

In the subsequent sections, we first silhouette different ways in which students can be negatively exposed to the posting of others on MOOC forums and then explain possible interaction pathways through which this exposure can spread in the MOOC.

\section{Are you exposed to a conducive learning environment?}
MOOCs, generally having the option of free and open registration, grow in an unruly manner. Students may join a MOOC in any week after the course starts. This has a serious negative consequence. Student cohorts are overwhelmed with loads of discussion forum content already posted, when they join the course. We hypothesize that, if students are not exposed to useful, informative and good quality posts on arrival to MOOC forums, they will not be exposed to a healthy learning environment and will find difficulty in deriving true utility from the immense potential that these learning networks have to offer. This determining factor, coupled with other influential factors such as noise (advertisements, inappropriate and impolite content posting) or questions remaining unanswered (indicative of lack of peer support), might make them less excited and motivated to participate, which in turn will increase their dropout chances. 

As a fair proxy for exposure in absence of view data, prior work has used common thread posting as an indicator of being exposed to all posts in the corresponding thread \cite{WenA:14,WenB:14,Yang:14}. However, in presence of view data, we consider percentage of good quality posts that the students viewed on a particular MOOC forum or discussion thread as an index for positive exposure and vice versa for negative exposure. Consequently, we outline the following ways to measure post quality that students in MOOCs can be exposed to:
\begin{enumerate}
\itemsep-0.4em
\item {\bf Directly inferred from data} (Weak proxy for post quality, but easily generalizable)
\begin{enumerate}
\item {\bf Highly upvoted and downvoted posts}: On MOOC forums, learners can vote on posts, depending on its relevance to the question asked or if it fosters healthy discussions on the appropriate thread. 
\item {\bf Posts from highly reputed users}: MOOCs such as Coursera maintain a reputation forum, which are fundamentally designed to provide incentives to learners for their good participation. Technically, reputation points for students are calculated as the sum of square roots of votes across all forum contributions, as defined by \cite{Huang:14} 
\end{enumerate}
\item {\bf Indirectly inferred from data} (Effective way to measure post quality by learning linguistically rich extraction patterns from unannotated text (using hand coded annotation procedure on forum posts, followed by supervised learning methods), but less generalizable)
\begin{enumerate}
\item {\bf Posts that are on-content}: To ensure that discussion forums act as facilitators of knowledge flow in the MOOC network, we intuitively expect content focused posts to be a fairly strong proxy for post quality. The only disadvantage is that content overlap between courses might be extremely negligible in some cases, for example, between a MOOC course on psychology and machine learning. Therefore, on-content posts would have to be separately labeled for MOOCs belonging to different domains.
\item {\bf Posts that are on-conduct}: Just like other online learning communities (Internet relay chats, Question answer forums such as Stackoverflow etc), MOOCs like Coursera too outline certain forum guidelines\footnote{\url {http://help.coursera.org/customer/portal/articles/1220499-forum-code-of-conduct}} to make the forums welcoming, easy to use and beneficial for participating students. It is expected that students adhere to these diffused norms of MOOC discussion forums to create a healthy learning community, by posting appropriate content, being polite and sensitive to controversial topics, staying on topic and voting wisely. There is an interesting prior work which relates adherence of group norms in an open source online community to increased participation benefits, in terms of higher chances of response elicitation \cite{Jain:13}. However, unlike what we attempt to capture now, the approach in this paper does not track how these norms spread in the community.
\item {\bf Posts that indicate high learner motivation and cognitive engagement}: Such a methodology for post labeling has been devised by \cite{WenA:14}. Linguistic cues were developed for capturing motivation levels, while the level of language abstraction was used as a measure for cognitive engagement.
\item {\bf Posts that express positive sentiment towards the course}: Such an approach for post labeling has been taken by \cite{WenB:14} to capture behavioral and affective trends in students' posts and can potentially be used as an index for post quality. 
\end{enumerate}
\end{enumerate}

As an outcome of exposure to good or bad quality posts, we hypothesize that, if majority of students' posts in a week are downvoted, off-content or off-conduct, they are more likely to receive very few and unsatisfactory responses. Therefore, there is a high chance that such students will develop a feeling of alienation and not infuse well with the MOOC community, leading to attrition. To further intensify this feeling of ``lack of community involvement" and influence the decision-making processes of newer student cohorts not to stay in the course, we very well know from prior work that ``rich club" phenomenon prevails in these online learning communities. Only a central core of students engage in persistent interactions leaving others out \cite{Vaq:13,SinhaB:14}. 

For the contrasting outcome, if students' post are a)motivating, b)intended to help peers, c)align well with linguistic norms and practices of the discussion forum, they will receive positive responses and feedback, which will in turn boost confidence and very likely increase their engagement in the MOOC.

\section{Modeling student attrition dynamics based on negative forum exposure}
We quantify negative influence that spreads in the post reply MOOC discussion forum networks, analogous to the manner in which products, ideas, norms and behaviors diffuse in social networks \cite{Louni:14,Rod:14} or an infectious disease spreads through a susceptible population \cite{Eas:10}. Let us consider a model where MOOC learners can be in the following 3 states throughout their participation trajectory: 
\begin{itemize}

\item {\bf Susceptible} : Students who are vulnerable to dropout
\item {\bf Affected} : Students who can affect other students' dropout behavior
\item {\bf Removed} : Students who eventually dropout at some time point in the MOOC 
\end{itemize}

We abbreviate these three states by S, A, R respectively. Initially, some students are in `A' state (say, top `k' students who do not have good quality posting). Other students are in `S' state. Based on non-exposure to a conducive learning environment (lack of exposure to good quality posts), these students move to `A' state. After `t' time-steps (weeks), such students in `A' state eventually dropout and move to `R' state.

To explain the branching process by which negative influence might spread, consider the following:
\begin{itemize}
\item {\bf First wave}: Suppose that an affected (`A') student enters the MOOC forums, and negatively influences students who view his post independently with a probability of `p'. Further, suppose that he influences `k' of these susceptible (`S') students while he is affected; let's call these `k' students the first wave of propogation. Based on the `p' value, some of the students in the first wave may get affected (turn from S$\rightarrow$A), while others may not.
\item {\bf Second wave}:  Now, each student in the first wave participates in the MOOC forums and negatively influences `k' different students, resulting in a second wave of k*k = k$^2$ students. Each affected student in the first wave spreads the negative influence to each of the `k' second-wave students, again independently with probability `p'.
\item {\bf Subsequent waves}: Further waves are formed in the same way, by `k' new students getting exposed to each student in the current wave, and turning from S$\rightarrow$A independently with probability `p'.
\end{itemize}

It is important to note that this spread of negative influence can be aggressive or mild depending on the value of `p'. There are really only two possibilities for the negative influence in the branching process model as described above: it reaches a wave where it affects no student, thus dying out after a finite number of steps; or it continues to affect students in every wave, proceeding infinitely through the MOOC contact network. We can formalize a simple condition to tell these two possibilities apart. If `p*k' is $<$ 1 (the size of the negative influence spread is constantly trending downward), then with probability 1, the negative influence dies out after a finite number of waves. If `p*k' $>$ 1 (the size of negative influence spread is constantly trending upward), then with probability greater than 0, the negative influence persists by affecting atleast one student in each wave. Thus, to reduce the combined value of `p*k', either MOOC instructors could quarantine students to reduce the quantity `k' (by say, guiding or recommending good quality forum posts), or encourage better posting behavior to reduce the quantity `p' (by say, intervening politely when a students' post is off content/off conduct). 

\section{Caveats and Implications for MOOCs}
Although the contact network in MOOCs can be arbitrarily complex, we have outlined a basic formalization that captures how students' lack of exposure to good quality posts on MOOC discussion forums or negative exposure propogates through the MOOC network, in turn affecting students' participation behavior. However, it is important to mention the underlying assumptions behind formulating these equations: 
\begin{enumerate}
\item Probability of influence due to negative exposure `p' is same for all students (p$\neq$1)
\item Every susceptible student is exposed to every affected student (views the corresponding forum or thread). Looking through the lens of our model, prior work using survival analysis has intuitively assumed p=1 for all students \cite{WenA:14,WenB:14,Yang:14}, which is not the case with a real world and diverse online community such as a MOOC. Depending on closeness of contact reflected in MOOC social structure leading to contagion, we could assign a different `$p_{v,w}$' to each pair of students {\em v} and {\em w} for which {\em v} links to {\em w} in the directed MOOC network. Here, higher values of `$p_{v,w}$' correspond to closer contact and more likely contagion, while lower values indicate less intensive contact.
\item The `p' value remains constant over `t' time steps, while the student is in state `A', which might not be very close to real world behavior exhibited in the MOOC. As the post becomes older (weeks pass on), less students are likely to view and get exposed to that content. So, `p' would be comparatively higher in the first few time steps since getting affected, than in later stages. Prior work has also shown that new student cohorts engage with just the past few weeks in the discussion forum (one to two), and not with prior weeks \cite{SinhaB:14}.
\end{enumerate}

\section {Conclusion}
In this work, we made an attempt to understand the dynamics of negative exposure that might influence attrition of students in MOOCs over time. Because the MOOC audience comprises of learners with diverse demographics, skillsets and learning preferences, understanding impact of students' peers on their exhibited behavior and interaction footprint is crucial for designing ways to maintain a conducive learning environment in MOOCs. The alternate perspective outlined in this work provides a lens into how interactions among students could possibly affect attrition. 

This will help course instructors a)in moving closer to the finer structure of the MOOC learning community and looking at how students are influenced by their particular network neighbors, rather than viewing the network as a relatively amorphous population of individuals and looking at effects in aggregate, b)in deciding which student communities be intervened depending on type of influence students have on their peers, c)in grouping students into conflict-free teams for effectively accomplishing course related tasks in the MOOC \cite{SinhaA:14}.


\begin{thebibliography}{}

\bibitem[\protect\citename{Belanger and Jessica}2013]{Belanger:13}
Belanger, Y., and Jessica T. 2013. ``Bioelectricity: A Quantitative Approach Duke University's First MOOC"

\bibitem[\protect\citename{Dishion et al.}1994]{Dishion:94}
Dishion, T. J., Patterson, G. R., \& Griesler, P. C. 1994. ``Peer adaptations in the development
of antisocial behavior: A confluence model". {\em In L. R. Huesmann (Ed.), Aggressive
behavior: Current perspectives} (pp. 61–95). New York: Plenum

\bibitem[\protect\citename{Easley and Kleinberg}2010]{Eas:10}
Easley, D., \& Kleinberg, J. 2010. ``Networks, crowds, and markets: Reasoning about a highly connected world, Chapter 21". {\em Cambridge University Press}

\bibitem[\protect\citename{Esbensen and Deschenes}1998]{Esb:98}
Esbensen, F. A., \& Deschenes, E. P. 1998. ``A multisite examination of youth gang
membership: Does gender matter?". {\em Criminology}, 36, 799–827.

\bibitem[\protect\citename{Huang et al.}2014]{Huang:14}
Huang, J., Anirban D., Arpita G., Jane M., and Marc S. 2014. ``Superposter behavior in MOOC forums". {\em ACM Learing at Scale(L@S)}

\bibitem[\protect\citename{Jain et al.}2013]{Jain:13}
Jain, S., Sinha, T., Shah, A., Sharma, C., \& Rose, C. P. 2013. ``Impact of Group Norms in Eliciting Response in a Goal Driven Virtual Community". {\em In proceedings of 21st International Conference on Computers in Education (ICCE)}

\bibitem[\protect\citename{Kaplan et al.}1987]{Kaplan:87}
Kaplan, H. B., Johnson, R. J., \& Bailey, C. A. 1987. ``Deviant peers and deviant behavior:
Further elaboration of a model". {\em Social Psychology Quarterly}, 50, 277–284


\bibitem[\protect\citename{Loeber and Dishion}1987]{Loe:87}
Loeber, R., \& Dishion, T. J. 1987. ``Antisocial and delinquent youths: Methods for their
early identification". {\em In J. D. Burchard \& S. Burchard (Eds.), Prevention of delinquent
behavior} (pp. 75–89). Thousand Oaks, CA: Sage

\bibitem[\protect\citename{Louni and Subbalakshmi}2014]{Louni:14}
Louni, A., \& Subbalakshmi, K. P. 2014. ``Diffusion of Information in Social Networks". {\em In Social Networking} (pp. 1-22). Springer International Publishing.

\bibitem[\protect\citename{Mason and Watts}2012]{Mason:12}
Mason, W., \& Watts, D. J. 2012. ``Collaborative learning in networks". {\em Proceedings of the National Academy of Sciences} 109(3), 764-769.

\bibitem[\protect\citename{Rodriguez et al.}2014]{Rod:14}
Rodriguez, M. G., Leskovec, J., Balduzzi, D., \& Scholkopf, B. 2014. ``Uncovering the structure and temporal dynamics of information propagation". {\em Network Science}, 2(01), 26-65.

\bibitem[\protect\citename{Schmidt and Zach}2013]{Schmidt:13}
Schmidt, D. C., and Zach M. 2013. ``Producing and  Delivering a Coursera MOOC on Pattern-Oriented Software Architecture for Concurrent and Networked Software" 

\bibitem[\protect\citename{Sinha}2014a]{SinhaB:14}
Sinha, T. 2014a. ``Supporting MOOC Instruction with Social Network Analysis". {\em arXiv preprint arXiv:1401.5175} 

\bibitem[\protect\citename{Sinha}2014b]{SinhaA:14}
Sinha, T. 2014b. ``Together we stand, Together we fall, Together we win: Dynamic team formation in massive open online courses" {\em In Fifth International Conference on the Applications of Digital Information and Web Technologies (ICADIWT)} pp. 107-112. IEEE

\bibitem[\protect\citename{Sinha et al.}2014]{SinhaC:14}
Sinha, T., Li, N., Jermann, P., Dillenbourg, P. 2014. ``Capturing ``attrition intensifying" structural traits from didactic interaction sequences of MOOC learners". {\em Proceedings of the 2014 Empirical Methods in Natural Language Processing Workshop on Modeling Large Scale Social Interaction in Massively Open Online Courses}, Qatar, October, 2014.

\bibitem[\protect\citename{Vaquero and Cebrian}2013]{Vaq:13}
Vaquero, L. M., \& Cebrian, M. 2013. ``The rich club phenomenon in the classroom". {\em Scientific reports, 3}.

\bibitem[\protect\citename{Wen et al.}2014a]{WenA:14}
Wen, M., Yang, D., \& Rose, C. P. 2014a. ``Linguistic Reflections of Student Engagement in Massive Open Online Courses". {\em In Proceedings of the International Conference on 	Weblogs and Social Media} 

\bibitem[\protect\citename{Wen et al.}2014b]{WenB:14}
Wen, M., Yang, D., \& Rose, C. P. 2014b. ``Sentiment Analysis in MOOC Discussion Forums: What does it tell us?". {\em In Proceedings of Educational Data Mining}

\bibitem[\protect\citename{Yang et al.}2013]{Yang:13}
Yang, D., Sinha T., Adamson D., and Rose, C. P. 2013. ``Turn on, Tune in, Drop out: Anticipating student dropouts in Massive Open Online Courses" {\em In NIPS Workshop on Data Driven Education}

\bibitem[\protect\citename{Yang et al.}2014]{Yang:14}
Yang, D., Wen, M., \& Rose, C. P. 2014. ``Peer Influence on Attrition in Massive Open Online Courses"
\end{thebibliography}
\end{document}